# An Architecture for Context-Aware Knowledge Flow Management Systems

Ali Jarrahi[1] and Mohammad Reza Kangavari[2]

[1] Department of Computer Engineering, Iran University of Science and Technology
Tehran, Iran

[2] Department of Computer Engineering, Iran University of Science and Technology
Tehran, Iran

**Abstract**

The organizational knowledge is one of the most important and valuable assets of organizations. In such environment, organizations with broad, specialized and up-to-date knowledge, adequately using knowledge resources, will be more successful than their competitors. For effective use of knowledge, dynamic knowledge flow from the sources to destinations is essential. In this regard, a novel complex concept in knowledge management is the analysis, design and implementation of knowledge flow management systems. One of the major challenges in such systems is to explore the knowledge flow from the source to the recipient and control the flow for quality improvements concerning the users' needs as possible. Therefore, the purpose of this paper is to provide an architecture in order to solve this challenge. For this purpose, in addition to the architecture for knowledge flow management systems, a new node selection strategy is provided with higher success rate compared to previous strategies.

*Keywords: Knowledge Management, knowledge flow management system, knowledge sharing, knowledge quality improvement.*

## 1. Introduction

Nowadays, in knowledge era, the organizational knowledge is one of the most important and valuable assets of organizations [1]. Therefore, with the development of information technology, many organizations are becoming more intent on knowledge than labor [2]. In such environment, organizations with broad, specialized and up-to-date knowledge, adequately using knowledge resources will be more successful than their competitors.

For effective use of knowledge resources, they should be managed properly. The importance of this problem has redirected the focus of corporate managers from work process management to knowledge management [2]. For this reason, various experts from different fields such as management, economics, knowledge engineering, cognitive science and software engineering have come together to found a new field of science known as knowledge management (KM) [3]. The main idea of the KM is that organizations can effectively use the maximum capacity of existing knowledge inside and outside the organization in order to accomplish all their activities. Knowledge management refers to a set of processes that control the creation, dissemination and utilization of knowledge in order to achieve organizational objectives [4]. Therefore, effective knowledge management can enhance the creativity and competitiveness of knowledge-intensive teamwork [5].

Although organizational knowledge is one of the most important and valuable assets of organizations, solely maintaining great resources of knowledge does not necessarily benefits organizations, since knowledge is one of the few assets of which the value increases with use [1]. As result, dynamic knowledge flow from the sources to destinations is essential for effective use of knowledge [6]. Therefore, we can say that the final goal of knowledge management systems is the effective flow of knowledge and applying the transferred knowledge in the activities. Previous research on knowledge management mainly focused on organizational learning and on providing procedures and systems to encourage members to communicate, but seldom considered the efficiency and efficacy of knowledge sharing, especially the routing of knowledge in a geographically distributed team [5].

In this regard, a novel complex concept in knowledge management is the analysis, design and implementation of knowledge flow management systems. One of the major challenges in such systems is to explore the knowledge flow from the source to the recipient and control the flow for quality improvements concerning the users' needs as possible. Therefore, the purpose of this paper is to provide a knowledge flow management system (KFMS) architecture in order to solve this challenge. Accordingly, considering the user's context and needs in distributed organizations with specific topologies, the system chooses suitable routes to transfer knowledge from the sources in order to improve the quality of flowing knowledge, with the assumption that the nodes trust each other completely.

Experimental results have shown that using the provided system compared to traditional systems in retrieving explicit knowledge, the quality of activity results would be improved and the completion time would be reduced.

This paper is structured as follows. The related works are listed in section 2. Section 3 defines the related background information. Problem is defined in section 4. The KFMS is introduced in section 5. Section 6 introduces networking strategies. Experimental results are evaluated in section 7 and finally the conclusion and further works are provided in section 8.

## 2. Related Works

Semantic Web can be considered as a universal space of machinery intelligent computations in which all resources of knowledge are coupled together in a means-oriented manner, capable for understanding each other [7]. Tim Berners-Lee has said that the future Web, unlike current web, not only can be understood by humans but also can be processed and cognized by machines [7]. In other words, the Semantic Web is a network of information in global scale, in which the processing of knowledge is easily possible by machines. Semantic Web technologies provide advanced tools that can be used effectively in knowledge management systems. For instance, a knowledge management architecture for software development knowledge reuse is provided in [8] by using ontologies and Semantic Web technologies.

With the appearance and spread of mobile devices such as notebooks, PDA and smart phones, pervasive systems are becoming increasingly popular [9]. First, Mark Weiser in 1991 introduced term "pervasive" refers to the integration of devices into the users' daily life [9]. One field of pervasive computing is called context-aware systems. Context-aware systems are able to adapt their operations to the present context without explicit intervention of the user. "Context is any information that can be used to characterize the situation of an entity. An entity is a person, place, or object that is considered relevant to the interaction between a user and an application, including the user and applications themselves" [10]. Context information may be gained in different ways, such as using sensors, network information, device status, user profiles, and use of other sources [9].

A general architecture of knowledge flow management system is proposed [5]. The system organizes the knowledge flow network using workflow process patterns and the distribution of the knowledge energy in the team, passes knowledge from node to node, collects knowledge, processes data and case documents, tracks the contribution and use of knowledge, and assesses node energies.

## 3. Background

### 3.1 Knowledge Types and Knowledge Space

According to the multidimensional nature of knowledge, many researchers have classified organizational knowledge from different perspectives. From one perspective, knowledge is classified into three categories: tacit, explicit and implicit.

Basically, a class of knowledge is gained through experience, apprentice with a master and long talks with experts. This sort of knowledge is usually gained unconsciously after a long time. Most individual skills such as pottery, programming, etc are categorized in this class. This class of knowledge, that is hard to explicitly express, is called tacit knowledge [1, 11, 12].

Explicit knowledge is another type of knowledge that is much clearer than tacit knowledge and can be documented easily. This type of knowledge can be transmitted through formal languages, mathematical equations, logic, rules, procedures and symbols. It can be described and shared in the form of documents, manuals, programs and knowledge representation languages [1, 11, 12].

Nonaka has proposed the SECI model for organizational knowledge creation with four conversions between explicit knowledge and tacit knowledge as follows [12]:
- **Socialization (tacit to tacit):** is sharing tacit knowledge through face-to-face communication or shared experience. An example is an apprenticeship.
- **Externalization (tacit to explicit):** is developing concepts, which embed the combined tacit knowledge, and which enable its communication.
- **Combination (explicit to explicit):** is combination of various elements of explicit knowledge. An example is building a prototype.
- **Internalization (explicit to tacit):** is closely linked to learning by doing. The explicit knowledge becomes part of the individual's knowledge base (e.g. mental model) and becomes an asset for the organization.

In addition to tacit and explicit knowledge, some researchers consider another type of knowledge, called *implicit knowledge* [1, 13]. As defined by [13] "Implicit Knowledge is knowledge that can be externalized when needed but has not been externalized yet."

By putting together different dimensions of knowledge, knowledge space is generated. Different knowledge spaces are introduced by a subset of these dimensions, depending on the application [5, 14, 15, 16]. For instance, as shown in Fig. 1, a three dimensional knowledge space has introduced in [14].

Each point in the knowledge space, which is called a unit field [5], is associated with particular category of knowledge. Therefore, it is obvious that an organization's knowledge can be classified by providing a proper knowledge space.

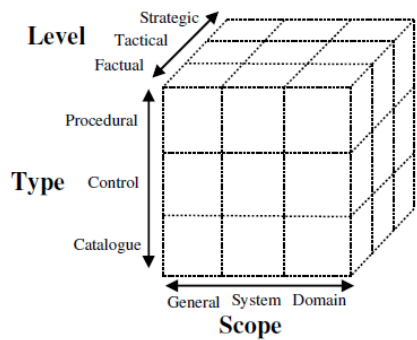

Fig. 1 An example of knowledge space [14].

3.2 Knowledge Flow and Its Elements

As defined by [16], knowledge flow (KF) is "A process of knowledge passing between people or knowledge processing mechanism."

Knowledge flow is a prerequisite and precondition for the application of knowledge (the ultimate goal of the KM). Since the only knowledge that is actively processed in the mind of an individual can be useful, the research on knowledge flow receives more and more attention [15]. According to above definition, knowledge flow consists of four key elements, as shown in Fig. 2.

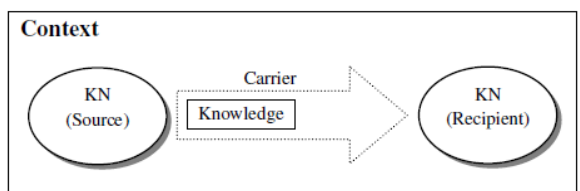

Fig. 2 Key elements of knowledge flow.

- *Knowledge Nodes:* knowledge sources (senders) together with recipients (customers) form the first element, namely Knowledge Nodes (KNs). A KN is possibly corresponded to a team member or an agent that is able to create, process and deliver knowledge. Senders and recipients of knowledge are also identified the direction of knowledge flows. In other words, the flow of knowledge can be the result of a knowledge source's trigger that is called Push strategy. Also, it can be prompted by a request for knowledge from the recipient that is called Pull strategy [1,15,16].
- *Knowledge:* The central element is knowledge that indicates the specific and sharable knowledge contents [1, 15, 16].
- *Carrier:* A carrier is the media that passes the knowledge content [16]. For instance, carriers can be based on local network, Internet or even magnetic tapes.
- *Context:* context represents the application environment wherein knowledge flow takes place. Knowledge flow without a common shared context between KNs cannot occur. In other words, it is necessary that both the source and the recipient share a common understanding of knowledge for a successful knowledge flow [1, 15].

In addition to the key elements listed above, some factors that greatly affect the occurrence of a successful knowledge flow are listed as follows.
- *Trust:* One of the most important and influential factors in the efficiency of knowledge interchange in an organization is the trust between KNs. Therefore, with the higher trust between KNs, the flow of knowledge between them will be more efficient and vice versa [17]. A trust model for knowledge flow networks has presented based on knowledge interaction experiences [17].
- *Knowledge Energy:* Knowledge energy is a parameter that expresses the level of a KN's knowledge and a person's cognitive and creative abilities in a unit field [5]. Hence, a KN with higher knowledge energy is more able to learn, use and create knowledge in its relevant field. Therefore, knowledge flow between two KNs only occurs when they have different knowledge energies in at least one unit field [5] (from the KN with higher knowledge energy to the KN with less knowledge energy).
A KN's knowledge energy in a single unit field is estimated by the degree of related knowledge in the KN, which is variable over time. A method for knowledge energy assessment of the KNs has presented using experts [5]. The assessments are done for all KNs and unit fields.

Although there is no limitation depicted in the definition of knowledge flow, a good knowledge flow is considered to be a flow of knowledge that delivers right knowledge to the right person at the right time.

Including a set of KNs, a knowledge flow network (KFN) is constituted by passing a knowledge flow through team members during a teamwork process. The KFN reflects the knowledge flows between KNs [2, 5]. By properly designing the knowledge flow network and controlling its execution process, the performance of the team will be dramatically increased [5].

## 4. Problem Definition

Basically, there is a considerable time since the required knowledge of a KN is identified until the knowledge must be delivered actually; especially if the identification is basis on work context information provided by the project management system. We believe that quality of the found explicit knowledge can be improved concerning the users' needs in the time constraint, using tacit and implicit knowledge of the related experts. The aim of the knowledge quality is level of relevance and effectiveness of knowledge to the activity or the requirement of the recipient. In this regard, this paper offers the KFMS architecture for organizations with distributed knowledge base, so the right knowledge would be delivered to the right person at right time. Right time is when the recipient needs the knowledge to perform the activity, which usually is a time interval.

If knowledge (even knowledge with the best quality) is not provided timely, it cannot be useful and effective to perform the desired activity. Consequently, the right time is preferred than the right knowledge in the KFMS architecture. Therefore, the KFMS tries to improve the quality of knowledge during the specified time constraint through finding a suitable path for the flow as possible. However, some assumptions are made for simplicity and complexity reduction of the problem listed below.

- *Network stability:* the network of organization is assumed as a stable network along with fixed topology. In other words, the desired network is not necessarily scalable and communication between nodes will never disconnect.
- *KNs Activity:* in this problem, it is assumed that the KNs are always active.
- *Mutual trust:* As noted, the Trust is one of the important factors affecting the flow of knowledge between the KNs. In this problem, it is assumed that all KNs trust each other completely. As a result, they offer their knowledge without any concerns. Also, there is no sabotage and prejudice in the knowledge processing.

## 5. KFMS Architecture

The KFMS system works as a platform for managing knowledge flows that will improve the quality of flowing knowledge concerning to the KNs' activity. For this purpose, the system effectively uses ontologies in order to manage and retrieve explicit knowledge, KNs' context information to determine the right knowledge and the right time and KNs' tacit as well as implicit knowledge to improve the quality of flowing knowledge.

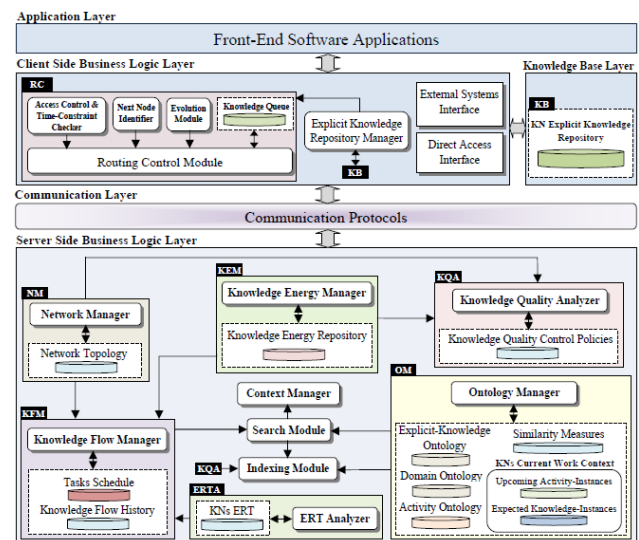

Fig. 3 The KFMS Architecture.

The KFMS architecture, a multi-layered client - server architecture, is presented in Fig. 3. In the Figure, if an arrow is drawn from module A to module B, it means that functions of module A are called by module B. The system can be structured in five layers: The server side business logic layer includes key elements of KFMS and is described in section 5-1. The communication layer comprises the communication protocols which are described in section 5-2. The knowledge base layer consists of explicit knowledge repository of KNs that contains original files of all types of organizational explicit knowledge resources which are associated with each KN. These sources include documents, best practices, and artifacts. The client side business logic layer includes additional elements for control knowledge flows as well as interfaces for communicate with elements of server side business logic layer which are described in section 5-3. The application layer includes all applications which are required. These applications can be classified as follows: *External Systems* such as project management system; The *User Tools* comprise a set of applications that KNs use for creating explicit knowledge and interaction with the system. These applications should have proper graphical user interface (GUI). Because a good GUI helps KNs for expressing their tacit/implicit knowledge (externalization), so then this knowledge can be knowledge of other KNs (internalization) [2]. The *Administrative Tools* consists of applications in order to interact with server side elements directly.

5.1 Server side business logic layer

The modules of the KFMS are divided in two parts, *server-side* and *client-side,* as shown in Fig. 3, to implement basic operations of the platform. This section is dedicated to server-side modules. In the next subsections, we describe in detail each of these modules.

5.1.1 Ontology Manager (OM)

The KFMS effectively uses ontologies in order to store and retrieve explicit knowledge. The ontologies in the platform will be represented using languages from the Semantic Web, namely RDF(S) and OWL [7, 18], and will be managed by Protégé framework [18] which is an open source Java framework for developing Semantic Web applications. This tool has been developed at Stanford University. The KFMS uses a set of ontologies as follows.

- **Domain Ontology:** The KFMS uses domain ontology, like SRS [8], to increase efficiency and accuracy of semantic searches in the system. The domain ontology is comprised of concepts and relations between them. It is required for understanding concepts in the organization. In other words, the organization's environment should be conceptualized in that concepts in the organization become understandable by the system. This ontology will be completed over time with completion of other ontologies. For this purpose, comprehensive ontologies, like WordNet [19], can be used effectively.
- **Activity Ontology:** One of properties of the KFMS is automatic recognition of the KNs' required knowledge for the Push knowledge flow. Therefore, it is necessary that the KFMS be aware of KNs' work context. For this purpose, it is required that all activities in the organization be conceptualized and be modeled as an ontology. In other words, different works that might be done by the KNs for their daily tasks have to be modeled and the relations between them are determined. These activities should be atoms, that couldn't be split into more sub-activities.
  There is a class in RDFS or OWL for each type of activity in the activity ontology, which describes the kind of activity in terms of required attributes. Also, possible relationships between various types of activities will be defined in the system and then the relations of all classes in the activity ontology will be identified in terms of these relations. An instance of the relevant class is created for every activity that is performed in the organization and the created instance is maintained in this sort of ontology.
- **Explicit-Knowledge Ontology:** The knowledge space of organization has to be defined carefully. There is a class per unit field in the knowledge space in the explicit-knowledge ontology, which describes that kind of knowledge in terms of required attributes. Also, possible relationships between various types of knowledge have to be defined in the system and then the relations of all classes in the explicit-knowledge ontology have to be identified in terms of these relations. An instance of the relevant class is created for explicit knowledge that is produced in the organization and the created instance is maintained in this ontology; but the relevant file will be stored in the explicit knowledge repository of the owner. *Link* is one of the attributes related to all classes of this ontology that refers to the exact location of that file.
  The attributes of the classes will be used to perform semantic searches. On the other hand, the explicit-knowledge ontology must be integrated with the activity ontology. In this case, the explicit knowledge that is related to the activities will be identified. When a KN is supposed to do "X" activity, the system would find most similar activity instance that somehow associated with the "X" activity, and then the explicit knowledge that was produced by the similar activity will be retrieved.

In addition to the ontologies, the OM also includes other repositories as follows.

- **Upcoming Activity-Instances:** All activities should be modeled in the activity ontology, as described before. Furthermore, when defining KNs' activities in a project management system, these activities have to be selected from the activity ontology, so that the system can be accurately informed KNs' upcoming activities. To do this, the project management system uses the relevant interface of ontology manager. Then these activities will be identified by context manager and will be provided for search module along with their time-constraints. After that, search module will provide a proper query using an ontology query language, such as SPARQL [20] or SQWRL [21], for search in the activity ontology by calling the relevant interface of ontology manager. Simultaneously, an instance of these activities will be created and maintained in this repository. By finishing the activities, relevant instances will be transmitted to the activity ontology for using in next similar activities.

- **Expected Knowledge-Instances:** new explicit knowledge has to be evaluated before it will be available to others in the system. Therefore, the respective instance will be placed temporarily in this repository after creating. Then, the instance will be transmitted to the explicit-knowledge ontology after final confirmation. Until this operation is not performed, the instance will not appear in search results.
- **Similarity Measures:** this repository contains information of the importance of each attribute of classes in the activity ontology and the explicit knowledge ontology. In other words, a weight is assigned to each attribute of classes that shows its importance. Obviously, the total weight of a class is equal to one. This information will be used by the ontology manager for ranking search results.

5.1.2 Context Manager

The KFMS should be aware of new activities ahead of each KN for Push knowledge flow. For this purpose, context manger recognizes these new activities along with their time constraints by interaction with the external project management system. After the recognition, this module provides information about these new activities, which perfectly matches the one of the classes in the activity ontology, along with the time constraint and brief description that be specified in the XML format and forwards it to search module by calling the relevant interface. Time constraint is the remaining time period up to the right time's upper bound.

5.1.3 Search Module

The search module is responsible for various searches in the KFMS. Therefore, it provides required interface to search for both Pull and Push knowledge flow. In the Pull approach, two cases can be considered: 1) *Using Keywords*, in this case the domain ontology is used to search. The results are ranked according to the number of matched concepts in the search term. In this way, in addition to the search term, a field is provided to set a time constraint by the recipient. 2) *Advanced search*, in this way, the knowledge space is provided to user in order to the desired unit field would be identified. Then associated attributes of the class are provided in a suitable form in order to the user able to initialize these attributes to search. In addition to these attributes, there are a field for problem description and a field to specify the time constraint. The results are ranked using the similarity measures.

Search element will be received in the XML format, then relevant query in an ontology query languages, such as SPARQL and SQWRL will be provided by the search module. After that, the search module will proceed to semantic search by calling relevant interface of the ontology manger. Then, it will call pertinent interface of the knowledge flow management module (KFM) in order to apply necessary controls for the flow of the most relevant knowledge. For this purpose, this module provides required information to the KFM, such as the link of found explicit knowledge, the recipient ID, the unit field of the found knowledge in the knowledge space, the problem description and the time constraint.

5.1.4 Knowledge Quality Analyzer (KQA)

The KQA is responsible for supporting the task of evaluation and quality control of new knowledge created in the organization. It is composed of a repository of knowledge quality control policies and knowledge quality analyzer module.

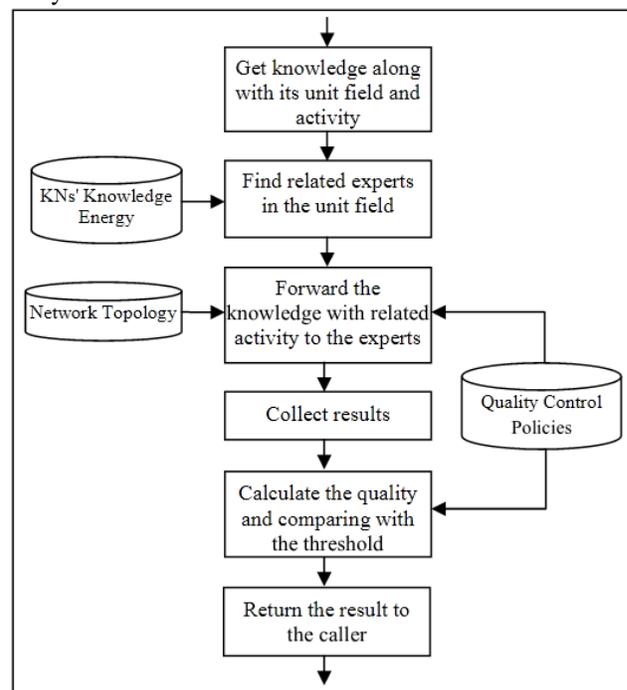

Fig. 4 Process of knowledge quality control.

The repository of knowledge quality control policies consists of information about quality assessment methods of new knowledge in any unit field. For example, a policy could be like this: "Post new knowledge for the three experts in the unit field associated with this new knowledge, collect results and calculate average and finally compare with the threshold."

The knowledge quality analyzer module is responsible for managing quality control process based on policies specified in the repository. For this purpose, it needs information about the knowledge energy of the KNs and the network topology, to determine the relevant experts and to locate the desired KNs, respectively. The quality control process is shown in Fig. 4.

### 5.1.5 Indexing Module

The indexing module is responsible for managing all matters pertaining to register new explicit knowledge in the system. For this purpose, the indexing module uses OM's interface to read ontologies and create instances, KQA's interface to evaluate the quality of new knowledge. The process of registering a knowledge element is presented in Fig. 5. In addition, this module can also perform the following tasks:
- Prevent duplicate registration of knowledge through interaction with search module.
- Support for multi-version files.

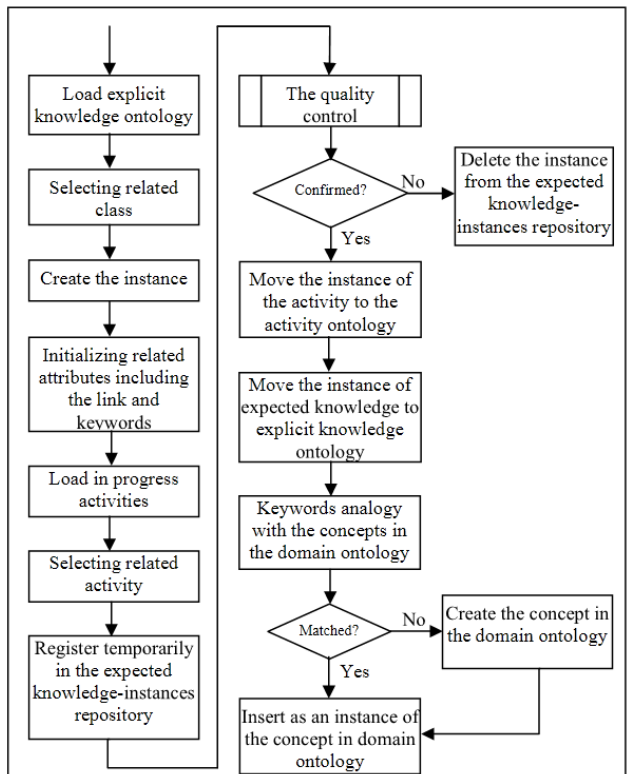

Fig. 5 Process of indexing new knowledge in the KFMS.

### 5.1.6 Network Manager (NM)

The NM is responsible for managing and maintaining information about network structure and links between the KNs. It provides proper interfaces to get information about network structure by other modules.

### 5.1.7 Knowledge Energy Manager (KEM)

The KEM is responsible for managing and maintaining information about the KNs' knowledge energy. It consists of repository for the knowledge energy and module of knowledge energy manager. The knowledge energy repository holds update information relating to state of the KNs' knowledge energy for each unit field in the knowledge space. A method of knowledge energy assessment in each unit field has presented using experts [5]. So, by evaluating and comparing KNs together, the relative rates per node per unit field will result in the form of a matrix. The assessment is repeated in the specified time period to be kept update.

The knowledge energy manager module is responsible for managing the knowledge energy repository. It provides information of KNs' knowledge energy for other modules by a proper interface.

### 5.1.8 Effective Response Time Analyzer (ERTA)

The ERTA is responsible for analyzing effective processing time of the KNs. This information is recorded and is updated in the KNs ERT repository for later use to determine the knowledge flow paths. The context data about each KN's processing status (the last response time) is gathered from the routing control module (on the client side business logic layer). Then, the effective response time for each KN is estimated using exponential moving average method.

### 5.1.9 Knowledge Flow Manager (KFM)

The KFM is responsible for managing knowledge flows in order to improve the quality of flowing knowledge considering the KNs' work context and the needs. It consists of a repository for knowledge flow history, a repository for tasks schedule and module of knowledge flow manager.

The knowledge flow history and the tasks schedule are used for awareness of the KNs' status (busy or idle) in different times. Tasks schedule information is obtained from the project management software. As a result, the busy time of each KN can be obtained in any time interval.

The knowledge flow manager module determines adequate flowing paths from the sources to the recipients using proper algorithm with the aim of improving quality of flowing knowledge. The node selection strategy, the time interval allocated to each KN and the order of knowledge provided to them must be dedicated with the purpose of defining adequate flowing paths. This algorithm uses the conscious strategy for networking. Different networking strategies are compared in the next section. The processing time interval is allocated to each KN using busy time information and their ERT. The following principle is used to order the knowledge providing.

***Principle 1:*** *Improving the quality of flowing knowledge on transition from a KN with less knowledge energy to a KN with more knowledge energy is more probable.*

The related algorithm is provided in Fig. 6. For this purpose, this module uses required information of KNs' knowledge energy, network topology and latest ERT of KNs by calling related interfaces of the KEM, NM and ERTA, respectively. An important point must be considered in the allocation of time interval to KNs. If new time interval has intersection with any different time interval of that KN, new interval must start after the start time of previous interval; otherwise, previous flow of knowledge would be disrupted.

```
Algorithm pathFinding
Create an empty list for the new knowledge flow (KFL);
Sort KNs DESC according to the unit field;
Remove KNs with equal or less knowledge energy than the recipient from the list;
For each KN_i in the list do
    Add to the KFL, first time period equal to ERT(KN_i), (KN_i, [t_1, t_1+ERT(KN_i)]),
    which [t_1, t_1+ERT(KN_i)]∩T_i=∅ by moving from t+TC to t;
For each time period, (KN_i, [t_i, t_j]), in the KFL do
    If (t_i-t_e(KN_i, t_i) < ERT(KN_i)) then
        t_i = max(t_s(KN_i, t_i), t_eKFL(t_i)) + ε;
    Else
        N = (t_i-t_e(KN_i, t_i))/ERT(KN_i);
        t_i = max(t_e(KN_i, t_i) + N*ERT(KN_i), t_eKFL(t_i));
Sort the KFL according to startTime;
Create flowing path using the KFL and the Network Topology;
Add the KFL to the knowledge flow history;
END Algorithm
---------------------------------------------------------------
t: Current Time
TC: Time Constraint
ERT(KN_i): Effective response time of KN_i
T_KF: Union intervals allocated in KFL
t_kp(KN_i): Set of knowledge processing intervals for KN_i
t_task(KN_i): Set of task intervals for KN_i
T_kp(KN_i): Union intervals in t_kp(KN_i)
T_task(KN_i): Union intervals in t_task(KN_i)
T_i : T_kp(KN_i) ∪ T_task(KN_i) ∪ T_KF
t_e(KN_i, t_x): Max(t, max EndTime before t_x in t_kp(KN_i) ∪ t_task(KN_i))
t_s(KN_i, t_x): Max(t, max StartTime before t_x in t_kp(KN_i) ∪ t_task(KN_i))
t_eKFL(t_x): Max(t, max EndTime before t_x in KFL)
```

Fig. 6 The path finding algorithm.

The KFM module provides proper interface in order to manage knowledge flows. During each call, the following information is received in the form of the parameters: recipient's ID, link of knowledge that should be flow, the unit field of flowing knowledge, the problem (description of problem or activity), and the time constraint. In response, as shown in Fig. 7, this module provides proper flowing path in specified format with the necessary control information such as access permission and the access time interval for each KN to improve quality along the path (no permission is equal to zero access time interval). The control message is sent to the source KN by calling the relevant interface of routing control module.

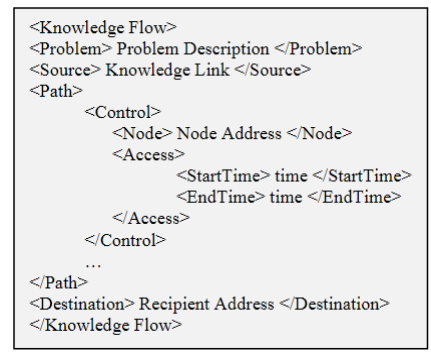

Fig. 7 The control message format.

### 5.2 Communication layer

Each module of the KFMS provides its capabilities to other modules through one or more interfaces. Modules communicate with each other by remote procedure call (RPC) or remote method invocation (RMI) mechanisms. However, more efficient and modern mechanisms such as web services and especially semantic web services can be used in order to increase performance and interoperability, particularly in Web-based implementation in Internet environment. In this case, the interfaces of each module will be implemented in the set of Semantic Web services.

The communication protocols are needed in order to interact between KNs in the network. The most important protocol is TCP / IP. On the other hand, the FTP protocol is also required to transfer related files of the explicit knowledge. The communication layer, in fact, consists of required protocols for interaction between modules of different network nodes.

### 5.3 Client side business logic layer

This layer consists of required modules for control knowledge flows according to the control message provided by the server side business logic layer and also the necessary interfaces for interaction of the applications with modules of the server side business logic layer that will be described in this section.

#### 5.3.1 Explicit knowledge repository manager

This module is responsible for access management to KN's explicit knowledge repository. In other words, all accesses to explicit knowledge resources are done through this module. By the same token, this module provides interfaces to access, read, write and edit files in the explicit knowledge repository to other modules and softwares.

It should be noted that by any change in stored knowledge, such as changing file location, this module calls the proper interface of ontology manager in order to update related instances.

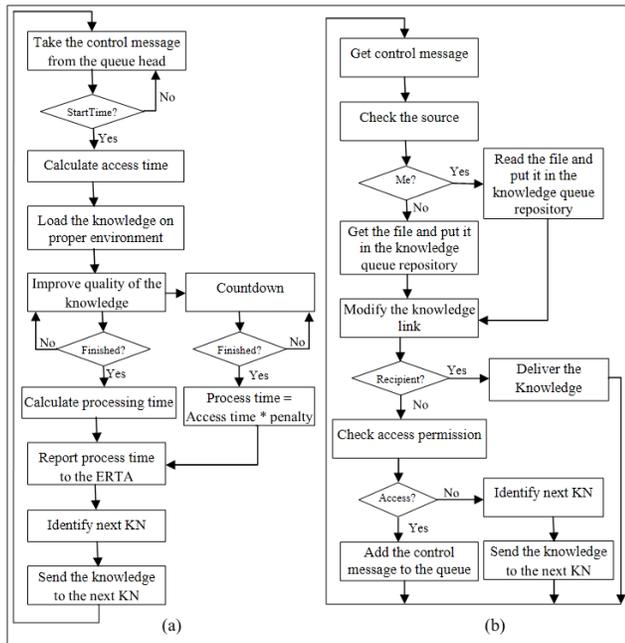

Fig. 8 The processes are controlled by the routing control module. (a) Process to improve the quality of knowledge available in the queue (b) Process to control new flowing knowledge.

### 5.3.2 Routing Control (RC)

The RC is responsible to control the knowledge flows and improvement process according to the control message. It includes components as follows.

- *Knowledge Queue:* This repository contains flowing knowledge files and related control messages. The control messages are sorted according to the start time.
- *Next Node Identifier:* This module, in each step, will determine the next KN according to the specified path in the control message.
- *Access Control & Time-Constraint Checker:* This module will control the access permission and process time interval using the control information (access time). If the KN fails to finish processing in this time interval, this module informs the routing control module to forward it to the next KN.
- *Evolution Module:* This module provides the knowledge to associated KN along with the problem description through user tools in the application layer in order to improve quality of knowledge considering the problem.

- *Routing Control Module:* This module is responsible for controlling process of knowledge flows using other modules of the RC. The routing control processes are shown in Fig. 8. These processes are performed concurrently.

### 5.3.3 Interfaces

The client side business logic layer also provides required interfaces for interaction between front-end applications and the server side business logic layer modules. These interfaces can be divided into two categories. One group provides interfaces for interaction of external systems with the KFMS, such as project management system, and second group provides interfaces for direct access of administrative tools to the server side modules. For example, the ontology manager interfaces to define ontologies and their classes.

## 6. Networking strategies

Flows of Knowledge between KNs can be established using different node selection strategies. Four strategies have introduced in [5] as follows.
1) *Random:* each node forwards the query to a randomly selected node.
2) *Greedy:* each node sends the query to a node with the highest energy.
3) *Generous:* each node forwards the query to a node with the lowest energy.
4) *Selfish:* a node sends the query to the node with higher energy. If the node has the top energy, it randomly selects a node with the same energy.

Here, we introduce *conscious* strategy that is used in the KFMS. In the conscious strategy, each node sends the query to an available node with the highest energy. The available node is a node that isn't too busy and can response in the specified time.

The impact of each strategy on the effectiveness of a knowledge flow is evaluated by the simulation done by [5].

A peer-to-peer knowledge flow network is constructed with 1000 autonomous KNs where KNs can query each other and knowledge flows through responses. The initial knowledge energies of the KNs are distributed randomly. 200 KNs are randomly selected that each of them creates a query per time slot. The maximum number of queries that can be accepted by a KN per time slot is supposed to be five. The results of interactions would react on the knowledge energies of respective KNs. The simulation runs for 10 time slots. Evaluation metrics are as follows.

1) *The number of successful interactions,* which represents the number of queries that have been responded by the KN successfully.
2) *The load of KNs,* which represents the number of queries that have been received by the KNs.
3) *The total successful, failure, and lost interactions.* A successful knowledge flow interaction relies on the KNs' knowledge energy difference and the risk factor. The lost interaction is an interaction that delayed over the maximum time slot.

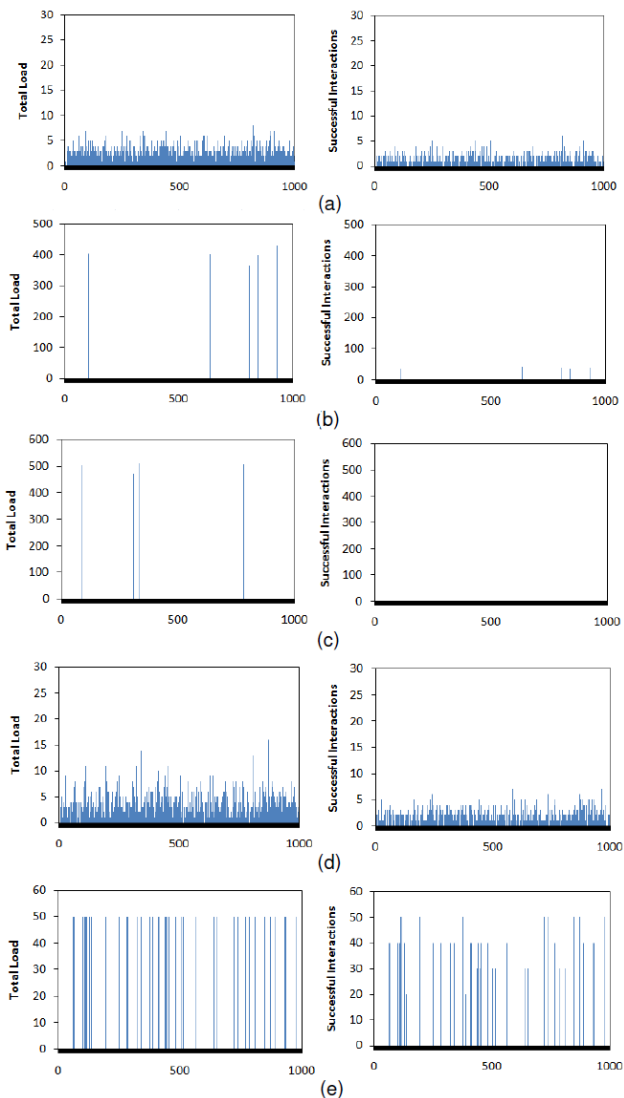

Fig. 9 The number of successful interactions and the loads of KNs. (a) Random, (b) Greedy, (c) Generous, (d) Selfish, (e) Conscious.

The results of the simulation are shown in Figs. 9 and 10. As can be observed, the conscious strategy shows the highest success rate. The loads are distributed between KNs with more knowledge energy and the distribution goes from the KNs with more knowledge energy to the KNs with less knowledge energy. Since the KNs are selected consciously, the success probability will be the highest; while there will be no lost interactions.

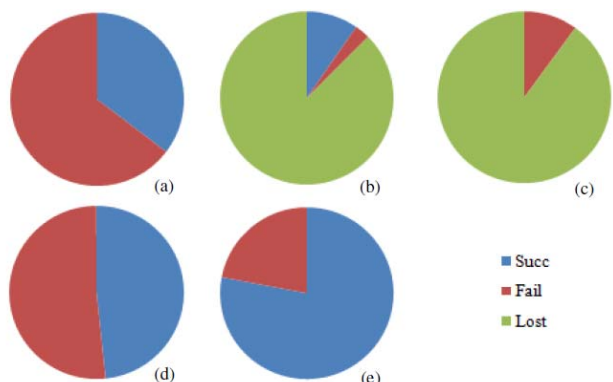

Fig. 10 Proportion of successful, failure, and lost interactions. (a) Random, (b) Greedy, (c) Generous, (d) Selfish, (e) Conscious.

## 7. Experimental results

In this section, effects of the KFMS on organizations performance will be assessed using an experiment in the case study. For this purpose, seven graduate and undergraduate students in computer engineering have been used. In this assessment, two different cases have been compared:

Case 1) A system is used only to retrieve the explicit knowledge, e.g. SRS [8].
Case 2) The KFMS is used.

Two undergraduate students, which had same knowledge energy, are used to do activities and the remaining members are used to quality improvement. In total, six activities are defined that are divided into three categories: two easy activities, two medium activities and two difficult activities. These activities have been performed in two different cases by the members (one member in case 1 and another in case 2). Results of the experiment are shown in Fig. 11 for effect on the completion time and the activity quality.

The results of the experiment have shown that using the provided system compared to traditional systems in retrieving explicit knowledge, the quality of activity results would be improved and the completion time would be reduced. However, the amount of improvement depends on activity's type and its difficulty level. This improvement is due to highest utilization of existing knowledge in the organization, including tacit and implicit knowledge of experts as well as providing the right knowledge in the beginning of each activity using work context information of KNs.

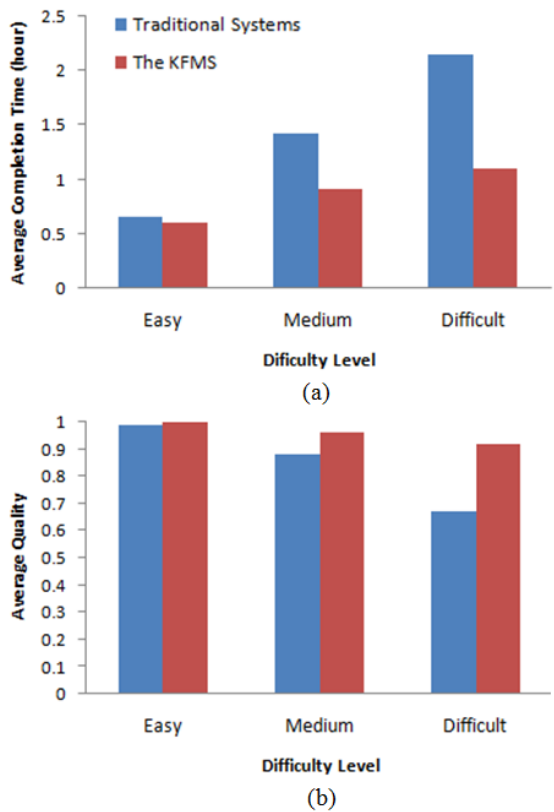

Fig. 11 Effects of the KFMS on organizations performance. (a) The completion time, (b) The activity quality.

## 8. Conclusions

Factors of competitive advantage include knowledge and the knowledge management [22]. In other words, organizations with broad, specialized and up-to-date knowledge, adequately using knowledge resources, will be more successful than their competitors. For effective use of knowledge, dynamic knowledge flow from the sources to destinations is essential. So, we can say that the final goal of knowledge management systems is the effective flow of knowledge and applying the transferred knowledge in the activities. Consequently, a novel complex concept in knowledge management is the analysis, design and implementation of knowledge flow management systems.

In this regard, this paper proposed the KFMS architecture. The main innovations of the proposed architecture are taking advantage of ontologies to retrieve explicit knowledge, quality improvement of flowing knowledge by dynamic knowledge flow network generation, supporting both Push and Pull knowledge flow and context-awareness of the KNs to provide the right knowledge at the right time. The process of knowledge flow in the KFSM architecture is summarized in Fig. 12. In addition to the architecture for knowledge flow management systems, the conscious networking strategy is provided with higher success rate compared to previous strategies.

The proposed architecture is prototyped for a software development team as case study. Experimental results have shown that using the provided system compared to traditional systems in retrieving explicit knowledge, the quality of activity results would be improved and the completion time would be reduced. However, the amount of improvement depends on activity's type and its difficulty level. This improvement is due to highest utilization of existing knowledge in the organization, including tacit and implicit knowledge of experts as well as providing the right knowledge in the beginning of each activity using work context information of KNs. Consequently, taking advantage of this system could be effective in success of organizations by the efficient use of organizational knowledge at the right time.

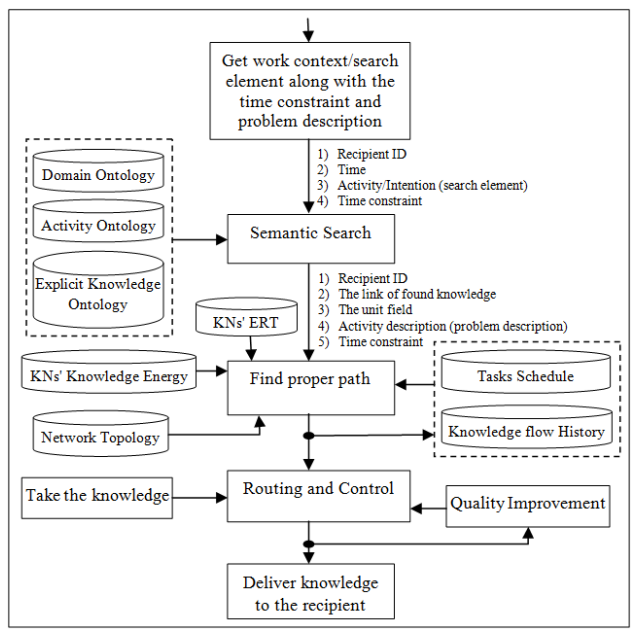

Fig. 12 The process of knowledge flow in the KFMS.

There are many factors that affect the success of knowledge flows. Some of these factors have not been considered in the KFMS architecture yet. Therefore, some future works are listed below.
- As noted, the trust is one of the factors that affect knowledge flows. Thus, trust should also be considered in routing algorithm of future KFMS.
- Required mechanisms should be used for the recipient satisfaction estimation. These mechanisms should assess the success of knowledge flow by monitoring the use of transmitted knowledge by the recipient and the system uses them to future flows of knowledge.
- Since in the proposed architecture, KNs are assumed always active, the mechanisms which analyze active patterns of KNs should be designed and applied.
- We are extending the KFMS architecture for peer-to-peer networks with dynamic topology. Preliminary results have been satisfactory.

**Ali Jarrahi** received his B.S. and M.S. degrees in computer engineering in 2008 and 2011, respectively. His research interests include knowledge management systems, data mining, and Algorithms & Theory.

**Dr. Mohammad Reza Kangavari** is Assistant Professor within Department of Computer Engineering (CE), Iran University of Science and Technology. His research interests include Artificial Intelligence, Data Mining, Knowledge Management, and Algorithms & Theory.